\documentclass[10pt,twocolumn]{IEEEtran}

\def\BibTeX{{\rm B\kern-.05em{\sc i\kern-.025em b}\kern-.08em
    T\kern-.1667em\lower.7ex\hbox{E}\kern-.125emX}}

\usepackage[dvipdf]{graphicx}
\usepackage{epsfig}
\usepackage{color}
\newtheorem{theorem}{Theorem}
\newtheorem{lemma}[theorem]{Lemma}
\usepackage{amsmath}
\usepackage{amssymb}
\begin{document}
\pagestyle{empty}
\thispagestyle{empty}
\title{\LARGE{Performance of Optimum and Suboptimum Combining Diversity Reception for Binary DPSK over Independent, Nonidentical Rayleigh Fading Channels}}

\author
{Hua Fu $\,$ and $\,$ Pooi Yuen Kam\\[-1pt]
Department of Electrical and Computer Engineering\\[-1pt]
National University of Singapore, Singapore 117576\\[-1pt]
e-mail: elefh@nus.edu.sg;$\,$ elekampy@nus.edu.sg\\[-1pt]
Tel: (65)6874-8892 $\;$  Fax: (65)6779-1103\\[-15pt]
}

\maketitle
\pagestyle{empty}
\thispagestyle{empty}
\begin{abstract}
This paper is concerned with the error performance analysis of binary differential phase shift keying 
with differential detection over the nonselective, Rayleigh fading channel with combining diversity 
reception. Space antenna diversity reception is assumed. The diversity branches are independent, but have nonidentically distributed statistics. 
The fading process in each branch is assumed to have an arbitrary Doppler spectrum with arbitrary Doppler bandwidth. Both optimum diversity reception and suboptimum diversity reception are considered. Results available previously apply only to the case of first and second-order diversity. Our results are more general in that the order of diversity is  arbitrary. Moreover, the bit error probability (BEP) result is obtained in an exact, closed-form expression which shows the behavior of the BEP as an explict function of the one-bit-interval fading correlation coefficient at the matched filter output, the mean signal-to-noise ratio per bit per branch and the order of diversity. A simple, more easily computable Chernoff bound to the BEP of the  optimum diversity detector is also derived.
\end{abstract}

%\begin{keywords}
%BEP, DPSK, Rayleigh fading, Chernoff bound, independent nonidentical channels, diversity reception.
%\end{keywords}

%-------------------------------------
\section{INTRODUCTION}
An efficient modulation technique for communication over a nonselective Rayleigh fading channel is differential phase shift keying (DPSK) with differential detection and combining diversity reception at the receiver [1]. 
Most previous analyses assume that the fading processes on the diversity branches are both independent and identically distributed (i.i.d.) [1]$-$[9]. However, in practical systems, this assumption is not always true [10]. The mean-square values of the fading attenuations are usually not identical for all diversity branches [11], [12], i.e., the average signal-to-noise ratio (SNR) of each branch is distinct. Therefore, it is of great practical interest and importance to study the performance of combining diversity reception for differentially detected DPSK over independent, nonidentical Rayleigh fading channels. With nonidentical fading channels, the combiner described in [7] and [8] is suboptimal, and it is well known that the BEP performance can be improved if each branch differential detector output is weighted, before combination, according to its channel condition [13]. We refer to this combining method with optimum weights for the branches as optimum combining reception. The bit error probability (BEP) performance of optimum combining reception was analyzed in [14]. However, the results are limited to the case of second-order diversity. Moreover, to the best of the authors' knowledge, as far as the {\em{exact, explict, closed-form}} BEP expression is concerned, there has not been much progress. The main contribution of this paper lies in providing the exact analysis of the BEP of binary DPSK with arbitrary order of diversity, both for the optimum combining reception and the suboptimum combining reception. The fading process is assumed here to have an arbitrary Doppler spectrum with arbitrary Doppler bandwidth. The BEP results here are derived in {\em{exact, closed-form}} expressions which depend {\em{explictly}} on the SNR per bit per branch, the fading correlation coefficient at the matched filter output over a bit interval, and the order of diversity. The solutions do not require any numerical integration for their actual evaluation. Moreover, a simple, tight, more easily computable Chernoff upper bound for the optimum diversity detector case is also obtained. 

The paper is organized as follows. In Section II, the system model and assumptions are introduced. The analyses for 
the BEP and the upper bound are given in Section III. Section IV presents numerical results. 
%****************************************************************************
\section{SYSTEM MODEL AND ASSUMPTIONS}%%% SYSTEM MODEL AND ASSUMPTIONS
With space antenna diversity reception over $L$ independent, nonidentical, frequency nonselective, Rayleigh fading 
branches with additive, white Gaussian noise (AWGN), the received signal over the $i$th branch, $i=1, 2, \cdots, L,$ during the $k$th bit interval $[kT, \; (k+1)T)$ is given, after matched filtering and sampling at time $t=(k+1)T$, 
by the statistic $\tilde{z}_{i}(k)$, where
\begin{equation}
\tilde{z}_{i}(k)=E_{b}^{1/2}e^{j\phi(k)}\tilde{a}_{i}(k)+\tilde{\nu}_{i}(k).
\end{equation}
The overhead $\sim$ denotes a complex quantity, and the superscript $*$ denotes the complex conjugate. $E_{b}$ is the energy per bit, $T$ is the bit duration, $\phi(k)$ is the data-modulated phase for the $k$th bit interval. The phase transition, $\Delta\phi(k)=\phi(k)-\phi(k-1)$, between the $k$th and the $(k-1)$th bit intervals carries the signal information. For binary DPSK, the data 0-bit is mapped onto the phase change $\Delta\phi(k)=0$, and the data 1-bit onto the phase change $\Delta\phi(k)=\pi$. We assume that the two data bits are equally likely. The multiplicative distortion $\tilde{a}_{i}(k)$ is given by
\begin{equation}
\tilde{a}_{i}(k)=\int_{kT}^{(k+1)T}\frac{\tilde{c}_{i}(t)}{T}\,d t.
\end{equation}
Here, $\left\{\tilde{c}_{i}(t)\right\}_{i=1}^{L}$ is a set of independent, nonidentically distributed, lowpass, 
complex Gaussian random processes with ${\rm{E}}\left[\tilde{c}_{i}(t)\right]=0$ and
$
{\rm{E}}[\tilde{c}_{i}(t)\tilde{c}_{i}^{*}(t-\tau)]=2R_{i}(\tau)
$
for each $i$. Each $\tilde{c}_{i}(t)$ represents the complex gain due to frequency nonselective Rayleigh fading of 
the $i$th branch. We assume that the spectrum of each $\tilde{c}_{i}(t)$ is even so that the in-phase component $Re\left[\tilde{c}_{i}(t)\right]$ and the quadrature phase component $Im\left[\tilde{c}_{i}(t)\right]$ are independent with the same covariance function $R_{i}(\tau)$. In (1), a rectangular data pulse shape $g(t)$, where $g(t)=1/\sqrt{T}$ for $0 \le t <T$ and zero elsewhere, is assumed so that each matched filter has a rectangular low-pass-equivalent impulse response $h_{i}(t)=g(T-t)$ for all $i$. This leads to the expression for $\tilde{a}_{i}(k)$ in (2). Thus, $\left\{\tilde{a}_{i}(k)\right\}_{k}$ is a sequence of zero-mean complex Gaussian random variables with covariance function 
\begin{eqnarray}
R_{i}(j) &=& \frac{1}{2}{\rm{E}}[\tilde{a}_{i}(k)\tilde{a}_{i}^{*}(k-j)]
\nonumber\\
&=& \frac{1}{T^{2}}\int_{kT}^{(k+1)T} \int_{(k-j)T}^{(k-j+1)T} R_{i}(t-\tau) d\tau dt.
\end{eqnarray}
The filtered noise $\tilde{\nu}_{i}(k)$ is given by
\begin{equation}
\tilde{\nu}_{i}(k)=\int_{kT}^{(k+1)T}\frac{\tilde{n}_{i}(t)}{\sqrt{T}}\,dt.
\end{equation}
Here, $\left\{\tilde{n}_{i}(t)\right\}_{i=1}^{L}$ is a set of i.i.d., lowpass, 
complex AWGN processes with ${\rm{E}}\left[\tilde{n}_{i}(t)\right]=0$ and
$
{\rm{E}}[\tilde{n}_{i}(t)\tilde{n}_{i}^{*}(t-\tau)]=N_{0} \delta(\tau)
$
so that $\left\{\tilde{\nu}_{i}(k)\right\}_{k}$ is a sequence of i.i.d., zero-mean, complex Gaussian variables with covariance function for each $i$
\begin{equation}
{\rm{E}}[\tilde{\nu}_{i}(k)\tilde{\nu}_{i}^{*}(j)]=N_{0}\, \delta_{kj}
\end{equation}
where $\delta_{kj}$ is Kronecker delta function.

For each $i \,$, $\tilde{c}_{i}(t)$ and $\tilde{n}_{i}(t)$ are mutually independent. 
For $i \ne j,$ $\; \{\tilde{c}_{i}(t),\tilde{n}_{i}(t)\}$ are independent of $\{\tilde{c}_{j}(t), \tilde{n}_{j}(t)\}$. 
The channel branches are nonidentical since the covariance function $R_{i}(\tau)$ depends on $i$.
For convenience of later application, the following parameters are defined. The quantity $\rho_{i}=\frac{R_{i}(1)}{R_{i}(0)}$ 
is the fading correlation coefficient at the matched filter output over an interval of $T$, and it is a measure 
of the fluctuation rate of the channel fading process; $\gamma_{i}=\frac{2E_{b}R_{i}(0)}{N_{0}}$ is the mean received 
SNR per bit over the $i$th branch. 
%***********************************************************************************************************
\section{THE PERFORMANCE ANALYSIS} %%% THE PERFORMANCE ANALYSIS
In this section, we will derive exact, closed-form BEP expressions and Chernoff bounds for differentially detected binary DPSK for the optimum combining diversity detector and the suboptimum combining diversity detector. In the Appendix, we show that the optimum combining detector with differential detection makes its decision on the $k$th transmitted bit using the likelihood ratio test
\begin{equation}
Re\left[\sum_{i=1}^{L} w_{i}\; \tilde{z}_{i}(k) \, \tilde{z}_{i}^{*}(k-1)\right]
\; {{{\rm{0\; bit}} \atop >} \atop {< \atop {\rm{1\; bit}}}} \;0
\end{equation}
where
\begin{equation}
w_{i}=\frac{\rho_i \gamma_i}{(1+\gamma_i)^2-(\rho_i \gamma_i)^2}.
\end{equation}
In the case of suboptimum combining diversity detection, one has $w_{i}=1$, for $i=1,2,\cdots,L$. Then the likelihood ratio test becomes
\begin{equation}
Re\left[\sum_{i=1}^{L} \tilde{z}_{i}(k) \, \tilde{z}_{i}^{*}(k-1)\right]
\; {{{\rm{0\; bit}} \atop >} \atop {< \atop {\rm{1\; bit}}}} \;0
\end{equation}
This is the same as the optimum combining diversity detection in the i.i.d. channel case.
\subsection{BEP analysis}%%%BEP and Chernoff bounds of 2DPSK
First, we consider the case of optimum reception. Because the BEP is the same whether a 0-bit or a 1-bit is sent, we assume that $\Delta \phi (k)=0$. Without loss of generality, it is assumed that $\phi (k)=\phi (k-1)=0$, and the BEP $P_{b}$ is given by
{\setlength\arraycolsep{-2pt}
\begin{eqnarray}
&&P_{b}= P\left\{{\rm{Re}}\left[\sum_{i=1}^{L} w_{i}\; \tilde{z}_{i}(k) \, \tilde{z}_{i}^{*}(k-1) \right]<0 
\Big{\vert}\Delta \phi (k)=0 \right\}
\\
&& =P\left[\sum_{i=1}^{L} w_{i} {\big{\vert}} \tilde{z}_{i}(k) + \tilde{z}_{i}(k-1) {\big{\vert}}^{2} < \sum_{i=1}^{L} w_{i} {\big{\vert}} \tilde{z}_{i}(k) - \tilde{z}_{i}(k-1) {\big{\vert}}^{2} \right].
\nonumber
\end{eqnarray}}
The alternative case where the 0-bit sent is due to $\phi (k)=\phi (k-1)=\pi$ gives an identical result. 
With $\phi (k)=\phi (k-1)=0$, one can see from (1) that for each $i=1, 2, \cdots, L$, $\tilde{z}_{i}(k)$ and $\tilde{z}_{i}(k-1)$ are both zero-mean complex Gaussian random variables (since each one is a sum of two zero-mean complex Gaussian random variables) with variances
\begin{equation}
{\rm{E}}[\vert \tilde{z}_{i}(k) \vert^2]={\rm{E}}[\vert \tilde{z}_{i}(k-1)\vert^2]=2E_b R_{i}(0)+N_0.
\end{equation}
Therefore, the quantity $\tilde{z}_{i}(k) + \tilde{z}_{i}(k-1)$ is complex Gaussian with mean zero and variance $4E_{b}R_{i}(0)+4E_{b}R_{i}(1)+2N_{0}$, and $\tilde{z}_{i}(k) - \tilde{z}_{i}(k-1)$ is complex Gaussian with mean zero and variance $4E_{b}R_{i}(0)-4E_{b}R_{i}(1)+2N_{0}$. Moreover, the quantities $\tilde{z}_{i}(k) + \tilde{z}_{i}(k-1)$
and $\tilde{z}_{i}(k) - \tilde{z}_{i}(k-1)$ are uncorrelated since
{\setlength\arraycolsep{0pt}
\begin{eqnarray}
&&{\rm{E}}\{[\tilde{z}_{i}(k)+\tilde{z}_{i}(k-1)]\,[\tilde{z}_{i}(k)-\tilde{z}_{i}(k-1)]^*\}={\rm{E}}[\vert \tilde{z}_{i}(k) \vert^2]
\nonumber\\
&&-{\rm{E}}[\vert \tilde{z}_{i}(k-1)\vert^2]-j2{\rm{E}}\{Im[\tilde{z}_{i}(k)\tilde{z}_{i}^{*}(k-1)]\}=0
\end{eqnarray}}
where $j=\sqrt{-1}$. The last step follows from (10) and the result
$
{\rm{E}}[\tilde{z}_{i}(k)\tilde{z}_{i}^{*}(k-1)]=2E_{b}R_{i}(1) 
$,
which is a purely real quantity. Thus, $\tilde{z}_{i}(k) + \tilde{z}_{i}(k-1)$ and $\tilde{z}_{i}(k)-\tilde{z}_{i}(k-1)$ are independent.
To proceed with computing (9), define
\begin{equation}
x_i =  w_{i} \vert \tilde{z}_{i}(k) + \tilde{z}_{i}(k-1) \vert^{2},
\quad
y_i = w_{i} \vert \tilde{z}_{i}(k) - \tilde{z}_{i}(k-1) \vert^{2}
\end{equation}
and
\begin{equation}
X = \sum_{i=1}^{L}\; x_i, 
\quad
Y = \sum_{i=1}^{L}\; y_i .
\end{equation}
Then the BEP $P_b$ in (9) can be rewritten as
\begin{equation}
P_b=P\left(X < Y \right).
\end{equation}
We note that each $x_i$ and each $y_i$ in (12) are chi-square-distributed with two degrees of freedom. The probability density function (PDF) can be evaluated as [1, (2.1-126)]
\begin{eqnarray}
p(x_i)&=&\frac{1}{w_i [4E_{s}R_{i}(0)+4E_{s}R_{i}(1)+2N_{0}]} \times
\\
&& {\rm{exp}}\left\{-
\frac{x_i}{w_i [4E_{s}R_{i}(0)+4E_{s}R_{i}(1)+2N_{0}]}\right\},
\nonumber\\
p(y_i)&=& \frac{1}{w_i [4E_{s}R_{i}(0)-4E_{s}R_{i}(1)+2N_{0}]} \times
\\
&& {\rm{exp}}\left\{-
\frac{y_i}{w_i [4E_{s}R_{i}(0)-4E_{s}R_{i}(1)+2N_{0}]}\right\}.
\nonumber
\end{eqnarray}
For each $i \,$, $x_{i}$ and $y_{i}$ are mutually independent, because $\tilde{z}_{i}(k) + \tilde{z}_{i}(k-1)$ is independent of $\tilde{z}_{i}(k) - \tilde{z}_{i}(k-1)$. Moreover, due to the independent channel assumption, for $i \ne j,$ $\; \{x_{i},y_{i}\}$ are independent of $\{x_{j},y_{j}\}$. Therefore, the quantity $X$ in (13) is independent of $Y$. 
The PDFs of $X$ and $Y$ can be obtained by using [1, (14.5-26)] and, after simplification, are given by
\begin{eqnarray}
p(X)&=&\sum_{i=1}^{L}\frac{A_i}{w_i [4E_{s}R_{i}(0)+4E_{s}R_{i}(1)+2N_{0}]}\times
\\
&&{\rm{exp}}\left[-
\frac{X}{w_i [4E_{s}R_{i}(0)+4E_{s}R_{i}(1)+2N_{0}]}\right],
\nonumber\\[5pt]
p(Y)&=&\sum_{i=1}^{L}\frac{B_i}{w_i [4E_{s}R_{i}(0)-4E_{s}R_{i}(1)+2N_{0}]}\times
\\
&&{\rm{exp}}\left[-\frac{Y}{w_i [4E_{s}R_{i}(0)-4E_{s}R_{i}(1)+2N_{0}]}\right]
\nonumber
\end{eqnarray}
where $A_i$ and $B_i$ are defined as
\begin{equation}
A_i = \prod_{m=1,m \ne i}^{L}\frac{\alpha_{i}}{\alpha_{i}-\alpha_{m}},
\quad
B_i = \prod_{n=1,n \ne i}^{L}\frac{\beta_{i}}{\beta_{i}-\beta_{n}}
\end{equation}
with
\begin{equation}
\alpha_i = \frac{\rho_i \gamma_i}{1+\gamma_i-\rho_i \gamma_i},
\quad
\beta_i = \frac{\rho_i \gamma_i}{1+\gamma_i+\rho_i \gamma_i}. 
\end{equation}
Finally, the BEP in (14) can be evaluated as
\begin{equation}
P_b=P\left(X < Y \right)=\int_{0}^{\infty}p(Y) \int_{0}^{Y}p(X)dX\, dY. 
\end{equation}
Using (17) and (18) in (21), the BEP is obtained as
\begin{equation}
P_b=\sum_{i=1}^{L} \sum_{j=1}^{L}\; A_{i}\,B_{j}\, \frac{\beta_j}{\alpha_{i}+\beta_{j}}.
\end{equation}
Note that (22) gives an exact, explicit, closed-form BEP expression. It depends only on the fading correlation coefficient $\rho_i$, the mean received SNR per bit per branch $\gamma_i$, and the number of diversity branches $L$. No numerical integration is needed for its actual evaluation. The quantities $A_i$ and $B_i$ are given in (19), where $\alpha_i$ and $\beta_i$ can be found in (20). 

Next, we consider the case of suboptimum reception. Again, it is assumed that $\phi (k)=\phi (k-1)=0$. The BEP of the suboptimum detector given in (8) can be obtained from the probability
%{\setlength\arraycolsep{-2pt}
\begin{equation}
P'_{b}=P\left[\sum_{i=1}^{L} \vert \tilde{z}_{i}(k) + \tilde{z}_{i}(k-1)\vert^{2} < \sum_{i=1}^{L} \vert \tilde{z}_{i}(k) - \tilde{z}_{i}(k-1)\vert^{2} \right].
\end{equation}%}
It is apparent that when the weights $w_i$$'$s satisfy $w_i=1$ for all $i=1, 2, \cdots, L$, the probability expression (23) is identical to that given in (9). Consequently, the BEP performance of suboptimum combining diversity reception is identical to that given in (22), where $A_i$ and $B_i$ are defined in (19). It can be shown that for suboptimum reception $\alpha_i$ and $\beta_i$ are given by
\begin{equation}
\alpha_i = 1+\gamma_i+\rho_i \gamma_i,
\quad
\beta_i = 1+\gamma_i-\rho_i \gamma_i. 
\end{equation}
Although exact and explicit, (22) is cumbersome and inconvenient to use when $L$ is large. Further, the behavior of the error probability (22) as a function of the various system parameters can not be easily seen. In the sequel, we will develop a simple, more easily computable Chernoff upper bound on the BEP.
\subsection{Chernoff bound analysis}%%%Chernoff bounds of 2DPSK
First, consider the optimum reception case. Applying the Chernoff bound [1, section 2.1.5] to (22) and noting that $X$ is independent of $Y$, we have
\begin{eqnarray}
P_b
 \leq  {\rm{E}} \left[e^{-s(X-Y)} \right]
= {\rm{E}} \left[e^{-sX} \right] {\rm{E}}\left[e^{sY} \right]
\end{eqnarray}
where $s \ge 0$ is the parameter to be optimized.
Using (12) and (13) in (25), and noting that for $i \ne j,$ $\; x_{i}$ is independent of $x_{j}$ and $y_{i}$ is independent of $y_{j}$, we have
\begin{eqnarray}
{\rm{E}} \left[e^{-sX} \right] 
&=& \prod_{i=1}^{L} {\rm{E}}\left[e^{-s\, w_{i} \vert \tilde{z}_{i}(k) + \tilde{z}_{i}(k-1) \vert^{2}} \right]
\\
{\rm{E}} \left[e^{sY} \right]
&=&\prod_{i=1}^{L} {\rm{E}}\left[e^{s\, w_{i} \vert \tilde{z}_{i}(k) - \tilde{z}_{i}(k-1) \vert^{2}} \right].
\end{eqnarray}
To proceed with computing (26) and (27), we use the following well-known lemma [15, (7.67)]

\begin{lemma}
If $x$ is a Gaussian random variable with mean $\nu$ and variance $\sigma^2$ and $\epsilon$ is any complex constant with real part less than $(2\sigma^2)^{-1}$, then
\end{lemma}
\begin{eqnarray}
{\rm{E}}\left[e^{\epsilon x^2} \right]=\frac{1}{\sqrt{1-2\epsilon\sigma^2}}\; e^{\epsilon\, \nu^2/(1-2\epsilon\sigma^2)}; \quad {\rm{Re}}(\epsilon)<\frac{1}{2\sigma^2}. \quad \blacksquare
\nonumber
\end{eqnarray}
Applying this lemma to (26) with $\epsilon= -s \, w_{i}$, and (27) with $\epsilon= s \, w_{i}$, and noting that
$\tilde{z}_{i}(k) + \tilde{z}_{i}(k-1)$ and $\tilde{z}_{i}(k) - \tilde{z}_{i}(k-1)$ are zero-mean complex Gaussian with variances $4E_{b}R_{i}(0)+4E_{b}R_{i}(1)+2N_{0}$ and $4E_{b}R_{i}(0)-4E_{b}R_{i}(1)+2N_{0}$, respectively, the Chernoff bound in (25) can be evaluated as
\begin{eqnarray}
&&{\rm{E}} \left[e^{-s X} \right]\; {\rm{E}}\left[e^{s Y} \right]
\\
&&=\prod_{i=1}^{L} \frac{1}{\left(1+ 4s \frac{\rho_i \gamma_i}{1+\gamma_i-\rho_i\gamma_i}N_{0}\right) 
\left(1- 4s \frac{\rho_i \gamma_i}{1+\gamma_i+\rho_i\gamma_i}N_{0}\right)}
\nonumber
\end{eqnarray}
where 
\begin{equation}
0< s <\frac{1}{4 \frac{\rho_i \gamma_i}{1+\gamma_i+\rho_i\gamma_i}N_{0}}. 
\end{equation}
The tighest upper bound is obtained by selecting $s$ that minimizes (28). This is equivalent to selecting $s$ that maximises the quantity $\prod\limits_{i=1}^{L} \left[\left(1+ 4s \frac{\rho_i \gamma_i} {1+\gamma_i- \rho_i\gamma_i} N_{0}\right) \left(1- 4s \frac{\rho_i \gamma_i}{1+\gamma_i+\rho_i\gamma_i}N_{0}\right)\right]$. The value $s$ that maximises the $i$th factor is determined by solving the equation
{\setlength\arraycolsep{0pt}
\begin{eqnarray}
&& \frac{d}{ds}\left[\left(1+4s 
\frac{\rho_i \gamma_i}{1+\gamma_i-\rho_i\gamma_i}N_{0}\right)
\left(1-4s 
\frac{\rho_i \gamma_i}{1+\gamma_i+\rho_i\gamma_i}N_{0}\right)  \right]
\nonumber\\[3pt]
&& \; =0.
\end{eqnarray}}
which gives the result
\begin{equation}
s = \frac{1}{4N_0}. 
\end{equation}
We note that this maximising value of $s$ is independent of index $i$ and falls within the allowable range of $s$ given in (29). 
Using (31) in (28), together with (25), one finally has
\begin{equation}
P_b \leq \prod_{i=1}^{L}\left[1-\left(\frac{\rho_i\gamma_i}{1+\gamma_i}\right)^2\right].
\end{equation}
The Chernoff bound in (32) can be tightened by a factor of 2 [16, section 4.2.4], that is
\begin{equation}
P_b \leq \frac{1}{2}\prod_{i=1}^{L}\left[1-\left(\frac{\rho_i\gamma_i}{1+\gamma_i}\right)^2\right]. \\[-7pt]
\end{equation}
For the case of i.i.d. channels, we have $\rho=\rho_i=\frac{R(1)}{R(0)}$ and $\gamma=\gamma_i=\frac{2E_{s}R(0)}{N_{0}}$ for each $i=1, 2, \cdots, L$. The improved Chernoff bound in (33) then reduces to 
\begin{equation}
P_b \leq \frac{1}{2}\left[1-\left(\frac{\rho\gamma}{1+\gamma}\right)^2\right]^L 
\end{equation}
which agrees with [7, eq.(4)]. Note that although the Chernoff bound (34) for the i.i.d. channel case can be deduced from (33) by setting $\rho=\rho_i$ and $\gamma=\gamma_i$, the BEP given in [7, eq.(3)] can not be obtained from (22). This is because for the i.i.d. channels the denominator terms of $A_i$ and $B_i$ in (19) are equal to zero.

Next, let us consider suboptimum reception. Although a nice, simple Chernoff bound to the error probability of the optimum detector can be obtained, computing the optimum bound parameter in the case of suboptimum detection turns out to be an analytically cumbrous problem, especially when the order of diversity is large. 
To illustrate this point, apply Chernoff bound to the error probability given in (23), we have
{\setlength\arraycolsep{0pt}
\begin{eqnarray}
&&P'_b \le {\rm{E}}\left[e^{-s' \sum\limits_{i=1}^{L} \vert \tilde{z}_{i}(k) + \tilde{z}_{i}(k-1)\vert^{2}}\right]
 {\rm{E}}\left[e^{s' \sum\limits_{i=1}^{L} \vert \tilde{z}_{i}(k) - \tilde{z}_{i}(k-1)\vert^{2} }\right]
\nonumber\\
&&=\prod_{i=1}^{L}\frac{1}{[1+ 4s'(1+\gamma_i +\rho_i \gamma_i)N_{0}]
[1- 4s'(1+\gamma_i -\rho_i \gamma_i)N_{0}]}
\nonumber\\
&&
\end{eqnarray}}
where 
\begin{equation}
0< s' <\frac{1}{4N_{0} (1+\gamma_i -\rho_i \gamma_i)}. 
\end{equation}   
The value $s'$ that minimizes the $i$th factor is determined as follows
\begin{equation}
\frac{d}{ds'}\{[1+ 4s'(1+\gamma_i +\rho_i \gamma_i)N_{0}][1- 4s'(1+\gamma_i -\rho_i \gamma_i)N_{0}] \}=0 
\end{equation}
which has solution given by
\begin{equation}
s' = \frac{\rho_i \gamma_i}{4N_{0} \left[(1+\gamma_i)^2 -(\rho_i \gamma_i)^2\right]}. 
\end{equation}   
Clearly, the local value $s'$ that minimizes each factor of (35) is dependent on index $i$, rather than equaling a constant independently of $i$. Therefore, the globally optimum bound parameter can only be obtained by solving
\begin{eqnarray}
&& \frac{d}{ds'} \prod_{i=1}^{L} {\Big{\{}}[1+ 4s'(1+\gamma_i +\rho_i \gamma_i)N_{0}] \times
\nonumber\\
&& \qquad  \qquad [1- 4s'(1+\gamma_i -\rho_i \gamma_i)N_{0}]{\Big{\}}}=0 
\end{eqnarray}
which is intractable when the order of diversity is large.
\section{NUMERICAL RESULTS}%%% NUMERICAL RESULTS
The BEP performance is plotted in Fig.1 and Fig.2 against the total average received SNR per bit. The order of diversity is set to $L=2$, and the fading correlation coefficient is set to $\rho=0.975$ for both the identical channel case and the nonidentical channel case (i.e., $\rho_1=\rho_2=0.975$). The abscissa represents the total average SNR per bit which is given by $\gamma_b=\gamma_1+\gamma_2$. The quantity $\eta$ is the fraction of the total average received bit energy devoted to diversity branch 1 ($1-\eta$ is devoted to diversity branch 2). For example, when $\eta=0.1$, $10\%$ of the average bit energy is devoted to branch 1, hence, $\gamma_1 = 0.1 \gamma_b$. Clearly, in the case of i.i.d. channels, $\gamma=\gamma_1 =\gamma_2=0.5\gamma_b$. The exact BEP for the nonidentical channel case is plotted using (22) and (19), together with (20) for the optimum diversity reception and (24) for the case of suboptimum diversity reception. The upper bound is plotted using (33) and (34) for nonidentical channel case and i.i.d. channel case, respectively. The exact BEP result for i.i.d. channel is computed using the work in [7, eq.(3a)]. 

From Fig.1 and Fig.2, several conclusions can be drawn. First, unequal SNR distribution among diversity branches degrades the BEP performance. For instance, when the total average SNR per bit $\gamma_b=15$ dB, the BEP is equal to $5.0234\times 10^{-3}$ for i.i.d. channel ($\gamma=\gamma_1=\gamma_2=12$ dB), $1.065\times 10^{-2}$ for optimum detection with $\eta=0.1$ ($\gamma_1=5$ dB and $\gamma_2=14.54$ dB) and $1.093 \times 10^{-2}$ for suboptimum detection with $\eta=0.1$. As $\eta$ increases to $0.5001$, we can see that the BEP performance of optimum and suboptimum detectors for nonidentical channels converges to the i.i.d. channels (see Fig.2). This observation also numerically validates our analytical result (22). Note that since the denominator terms of $A_i$ and $B_i$ in (19) are equal to zero for the case of i.i.d. channels, we have set $\eta=0.5001$ instead of using $\eta=0.5$ for our numerical investiagtion. 
Second, for the case of nonidentical channels, in comparison with the suboptimum detector given in (8), the optimum detector shown in (6) and (7) can substantially improve the BEP performance, especially in the regime of high total mean SNR. For example, for $\gamma_b=30$ dB and $\eta=0.1$, the BEP is $1.616\times 10^{-3}$ for suboptimum detection and is $6.710 \times 10^{-4}$ for optimum detection. Moreover, as the total average SNR goes to infinity, for the same order of diversity $L$ and the same value of correlation coefficient $\rho$, asymtotically the optimum detector for nonidentical channels performs identically to the combining detector for i.i.d. channels. Actually, if we let $\gamma_1 \rightarrow \infty$ and $\gamma_2 \rightarrow \infty$ in (33) and $\gamma \rightarrow \infty$ in (34), it can be easily seen that the BEP floor of the improved upper bound (33) for nonidentical channels is identical to that of (34) for i.i.d. channels. 
\appendices
\section{}
The task of the receiver is to determine from the received signals $\{\tilde{z}_{i}(k),\;\tilde{z}_{i}(k-1)\}_{i=1}^{L}$ 
which one of two possible values $0$ and $\pi$ of the phase difference $\Delta \phi (k)$ has maximum a posteriori probability (MAP). More precisely, the receiver will set $\Delta \phi (k)=\pi n$ whenever
\begin{equation}
P\left[\Delta \phi (k)=\pi\, m {\Big{\vert}} \left\{\tilde{z}_{i}(k),\;\tilde{z}_{i}(k-1)\right\}_{i=1}^{L}  \right], \;\; m=0, 1 
\end{equation}
is a maximum for $m=n$.

Assuming the two data bits are equally likely, MAP detection is equivalent to the maximum likelihood (ML) detection. That is, based on $\{\tilde{z}_{i}(k),\;\tilde{z}_{i}(k-1)\}_{i=1}^{L}$, we decide that $\Delta \phi (k)= \pi\, n$ whenever the PDF 
\begin{equation}
\Psi_m=p\left[\left\{\tilde{z}_{i}(k),\;\tilde{z}_{i}(k-1)\right\}_{i=1}^{L}{\Big{\vert}} \Delta \phi (k)=\pi\, m \right], \;\; m=0, 1 
\end{equation}
is a maximum for $m=n$. 
To simplify (41), we take the natural logarithm for both sides and use the independent channel assumption, resulting in log-likelihood
%{\setlength\arraycolsep{1pt}
\begin{eqnarray}
{\rm{log}}\Psi_m &=& \sum_{i=1}^{L} {\rm{log}}\left\{ p\left[\tilde{z}_{i}(k) {\Big{\vert}} \tilde{z}_{i}(k-1), \Delta \phi (k)= \pi\, m \right]\right\}
\nonumber\\[-3pt]
&& + \sum_{i=1}^{L}{\rm{log}}\left\{p\left[\tilde{z}_{i}(k-1) {\Big{\vert}} \Delta \phi (k)= \pi\, m \right] \right\}.
\end{eqnarray}%} 
Since the third term in (42) does not affect the log-likelihood decision, we only need consider the second term in the computation of the matrics. 
Conditioning on $\tilde{z}_{i}(k-1)$ and $\Delta \phi (k)= \pi\, m$, the quantity $\tilde{z}_{i}(k)$ is a Gaussian random variable with mean
$
\frac{2R_{i}(1)E_{b}}{2E_{b}R_{i}(0)+N_0} \tilde{z}_{i}(k-1)  e^{\pi\, m} 
$
and variance
$
2E_{b}R_{i}(0)+N_{0}-\frac{4E_{b}^{2}\, R_{i}^{2}(1)}{2E_{b}R_{i}(0)+N_{0}}
$ [7].
Using the conditional PDF of $\tilde{z}_{i}(k)$ in (42), and after manipulation and simplification, one has
\begin{eqnarray}
&& {\rm{log}}\Psi_m = C+ \frac{2}{N_0} \times
\nonumber\\[-5pt]
&& \quad Re\left[\sum_{i=1}^{L} \frac{\rho_i \gamma_i}{(1+\gamma_i)^2-(\rho_i \gamma_i)^2} \tilde{z}_{i}(k) \, \tilde{z}_{i}^{*}(k-1)e^{-j \pi\, m}\right]
\nonumber
\end{eqnarray}
where $C$ represents the constant term which does not affect the likelihood decision. The likelihood ratio test in (6) and (7) then follows.\\[-20pt]

%--------------------------------------------------------------------
\begin{figure}[t]
\center
\includegraphics[height=0.47\textwidth,width=0.5\textwidth]
{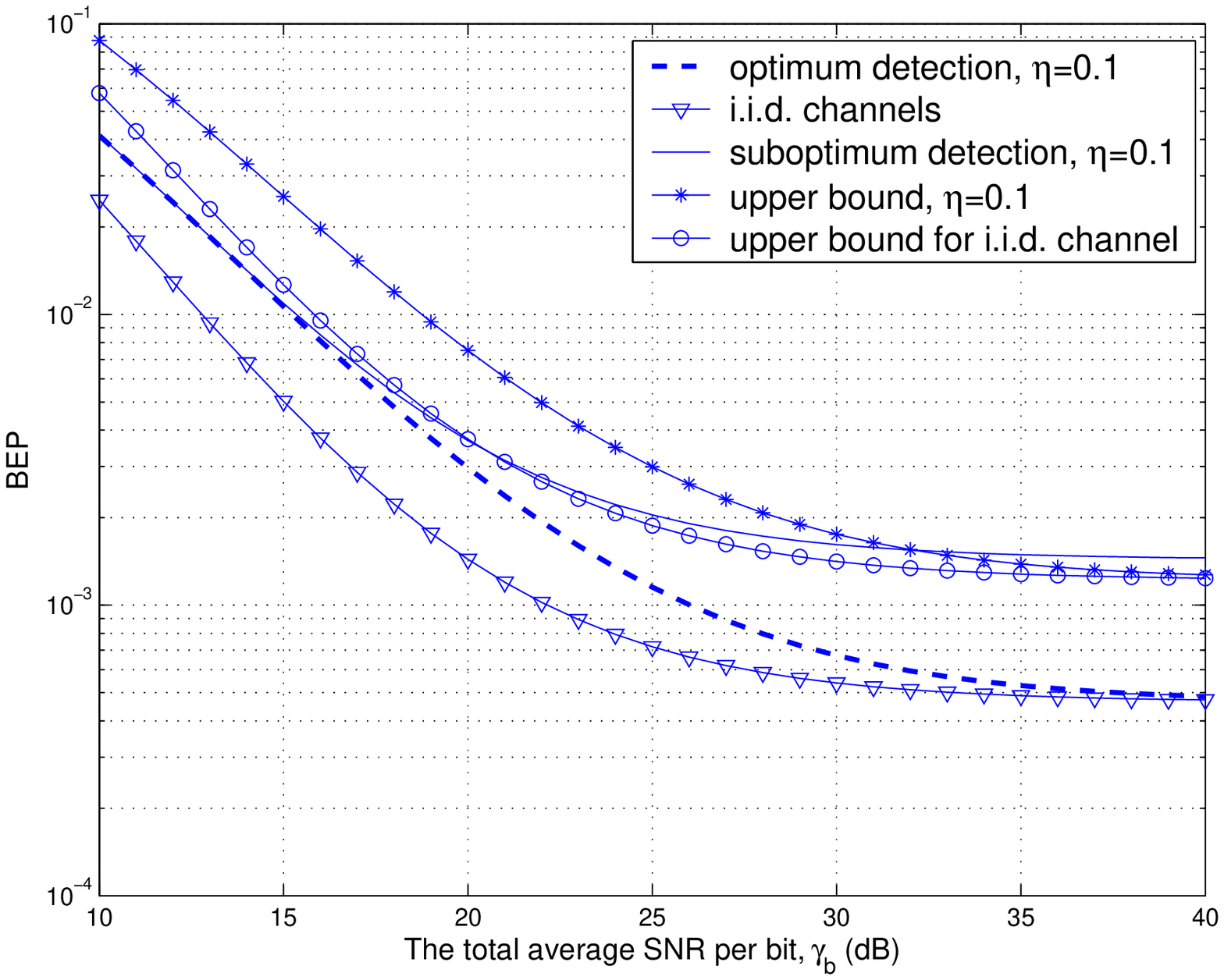}\\
\caption{BEP comparison for $L=2$; $\rho=0.975$.}
\end{figure}
%--------------------------------------------------------------------
\begin{figure}
\center
\includegraphics[height=0.47\textwidth,width=0.5\textwidth]
{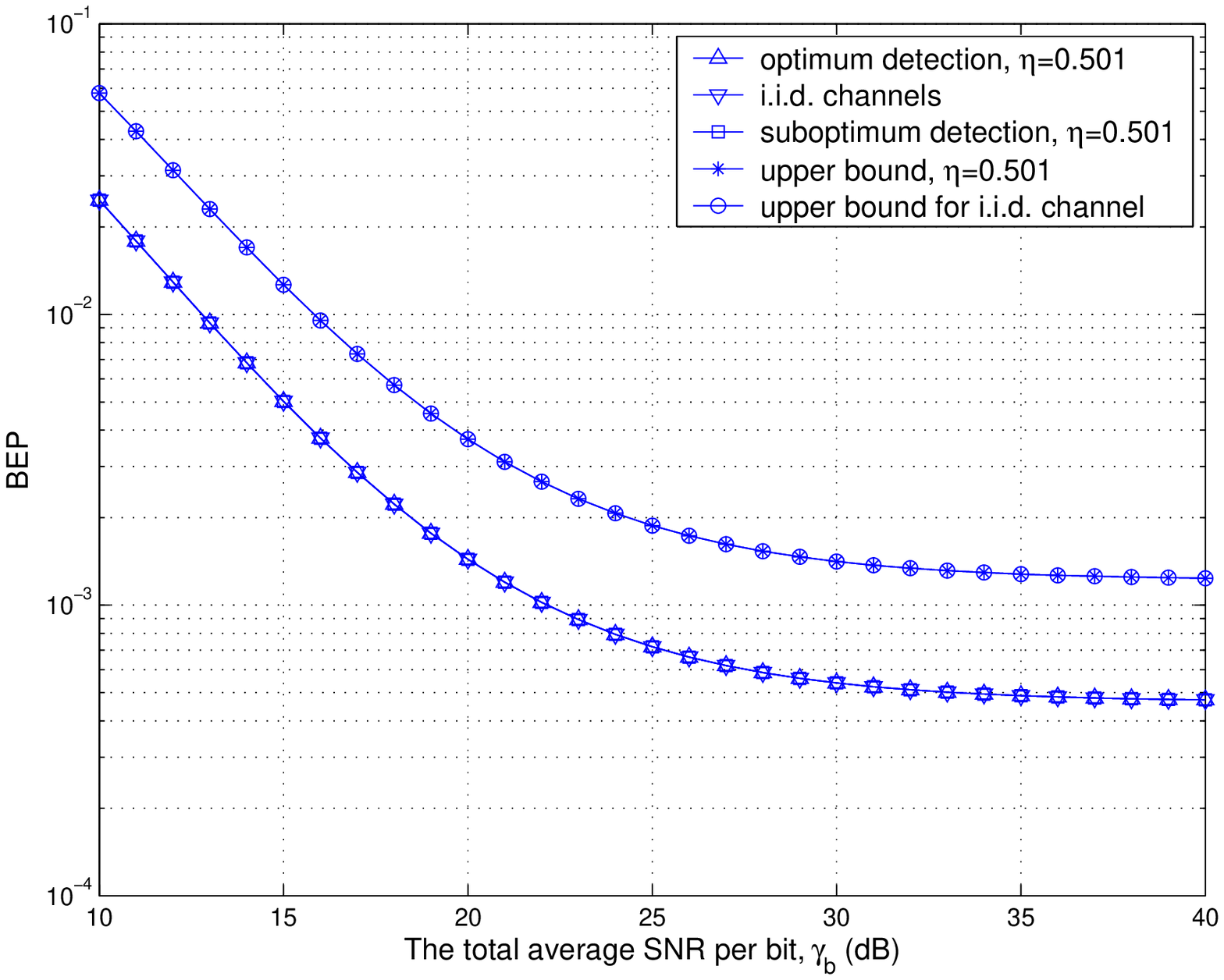}
\caption{BEP comparison for $L=2$; $\rho=0.975$.}
\end{figure}
%--------------------------------------------------------------------
%
\end{document}